\documentclass{SCIS2024}

\usepackage{amsmath} 
\usepackage{balance} 
\usepackage{booktabs} 
\usepackage{caption}
\usepackage{comment}
\usepackage{enumitem}
\setlist[itemize]{leftmargin=2.2em}

\usepackage{fancyvrb} 
\usepackage[symbol]{footmisc}
\usepackage{graphicx} 
\usepackage{makecell} 
\usepackage{multirow} 
\usepackage{soul}
\usepackage{xcolor}
\usepackage{xspace} 
\usepackage{bm}
\usepackage{utfsym}
\usepackage{cancel}
\usepackage{pifont}
\usepackage{tikz}
\usepackage{subcaption}
\usepackage{tcolorbox}

\usepackage{listings}
\usepackage{enumitem}
\setlist[itemize]{align=parleft,left=0pt..1em}
\lstdefinestyle{SQL}{
    language=SQL,
    basicstyle=\ttfamily,
    keywordstyle=\color{blue}\bfseries,
    commentstyle=\color{green!40!black},
    stringstyle=\color{red},
    showstringspaces=false,
    breaklines=true,
    xleftmargin=\parindent, 
    numbers=none, 
    captionpos=b, 
    columns=fullflexible 
}

\newcommand{\revision}[1]{{#1}}

\newcommand\dbname{\ensuremath{\textsf{NeurDB}}\xspace}
\newcommand\aixdb{AI$\times$DB\xspace}

\begin{document}

\setcounter{page}{1}


\title{\dbname: An AI-powered Autonomous Data System
}{\dbname: An AI-powered Autonomous Data System}

\author[1]{Beng Chin Ooi}{}
\author[1]{Shaofeng Cai}{}
\author[2]{Gang Chen}{}
\author[3]{Yanyan Shen}{}
\author[1]{Kian-Lee Tan}{}
\author[4]{Yuncheng Wu}{}
\author[1]{\\Xiaokui Xiao}{}
\author[1]{Naili Xing}{}
\author[1]{Cong Yue}{}
\author[1]{Lingze Zeng}{}
\author[5]{Meihui Zhang}{}
\author[1]{Zhanhao Zhao}{{zhanhao@nus.edu.sg}}

\AuthorMark{Beng Chin Ooi}

\AuthorCitation{Beng Chin Ooi, Shaofeng Cai, Gang Chen, Yanyan Shen, Kian-Lee Tan, Yuncheng Wu, Xiaokui Xiao, Naili Xing, Cong Yue, Lingze Zeng, Meihui Zhang, Zhaohao Zhao}



\address[1]{National University of Singapore, Singapore}
\address[2]{Zhejiang University, China}
\address[3]{Shanghai Jiao Tong University, China}
\address[4]{Renmin University of China, China}
\address[5]{Beijing Institute of Technology, China}

\abstract{
In the wake of rapid advancements in artificial intelligence (AI), we stand on the brink of a transformative leap in data systems.
The imminent fusion of AI and DB (\aixdb) promises a new generation of data systems, which will relieve the burden on end-users across all industry sectors by featuring AI-enhanced functionalities, such as personalized and automated in-database AI-powered analytics, self-driving capabilities for improved system performance, etc. 
In this paper, we explore the evolution of data systems with a focus
on deepening the fusion of AI and DB. 
\revision{We present \dbname, 
an AI-powered autonomous
data system designed to fully embrace AI design in each major system component and provide in-database AI-powered analytics.}
We outline the conceptual and architectural overview of \dbname, discuss its design choices and key components,
and report its current development and future plan.
}



\maketitle


\setcounter{section}{0}
\section{Introduction} \label{sec:introduction}


In tandem with digitization and digitalization initiatives across all industries, data systems have evolved to become the cornerstone of mainstream online applications integral to our everyday lives.
Traditionally, data systems (aka databases or data management systems) have been designed to meet the transactional and analytical data processing requirements of these applications.
However, with the rapid advancements in artificial intelligence (AI), we are witnessing a surge in AI-powered applications across sectors, which aim to relieve the burden on end-users by offering users personalized, automated AI-enhanced analytics and predictions.
To support such emerging AI applications, we are now standing on the brink of a transformative leap in data systems, poised to realize a data system that genuinely aligns with the demand of end-users for intelligent and automatic data services.



\begin{figure*}[t]
\centering
\includegraphics[width=0.85\linewidth]{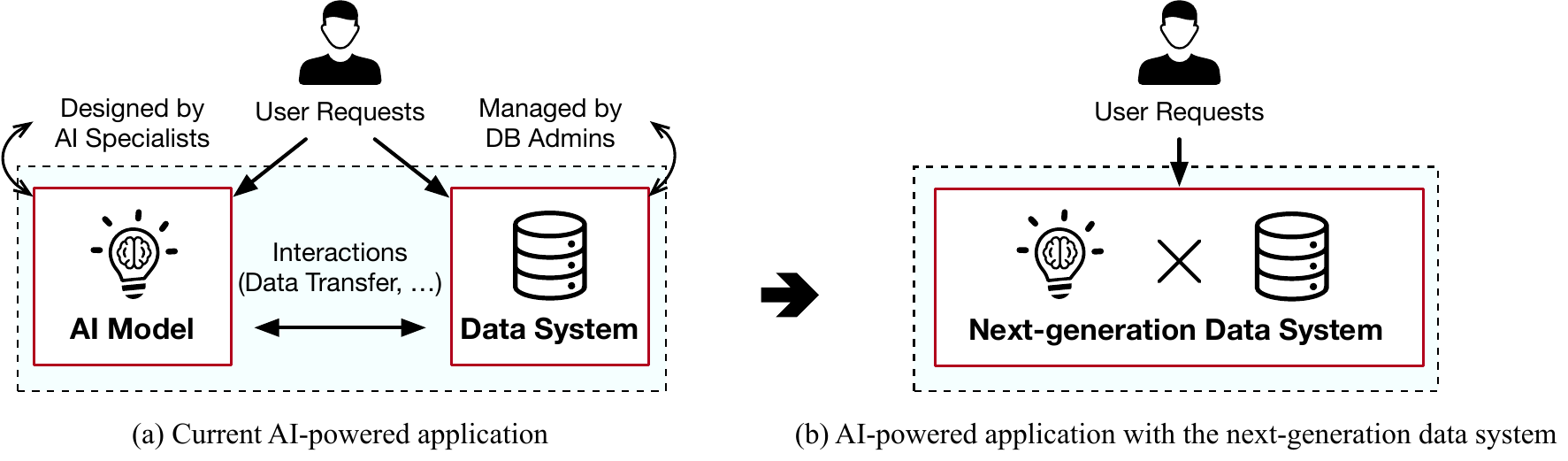}
\caption{Evolution to AI-powered applications with next-generation data systems}
\label{fig:intro}
\end{figure*}

At present, the realms of data systems and AI models exist as independent fields, each fundamental in its own right but isolated in functionality.
Take the healthcare sector as an example: clinicians require AI-powered assistance for disease progression modeling, and their data scientists need to navigate the complexities of both AI and database systems to construct 
prediction tasks.
This separation presents significant challenges in the development and operational efficiency of AI applications, as illustrated in Figure~\ref{fig:intro}(a).
The need for expertise in both domains to create and integrate these systems adds complexity to application development.
Further, user requests often trigger extensive interactions between AI models and databases, whether for data retrieval, model training, or fine-tuning, thereby impacting application performance.
These barriers hinder the user-friendliness and operational efficiency of applications, posing a major obstacle to the broader adoption of AI technologies.




Next-generation data systems are set to experience a transformative revolution with the imminent fusion of AI and DB (\aixdb) to dismantle the barriers preventing end-users from fully embracing AI technologies.
As we envisioned in 2016~\cite{sigmodRecord}, AI and databases can mutually enhance each other’s capabilities.
Yet years later, despite considerable progress in integrating AI with data systems, a significant gap remains between the potential of this integration and its current state of usability.
As shown in Figure~\ref{fig:intro}(b), bridging this gap requires a seamless integration of AI and database technologies, creating a unified data system that transcends the capabilities of each in isolation.
The transition to this new technological paradigm is an ambitious journey.
With an expected influx in the data volume, continuous shifts in data dynamics and workloads, and the emergence of increasingly more complex AI models and applications, we anticipate a substantial endeavor in the realization of next-generation data systems.
Such evolution promises not only to upgrade system functionalities through in-database AI-powered analytics but also to spur the development of novel AI-driven data system architectures and methodologies.



In this position paper, we trace the evolutionary trajectory of data systems and characterize related works on AI$\times$DB, establishing the timeliness and necessity of advancing toward next-generation AI-powered autonomous data systems.
We showcase promising AI-powered applications for which we believe the next-generation data system has much to offer.
Through a comprehensive analysis of user needs derived from these applications, we identify key challenges in deepening the fusion of AI and DB to realize next-generation data systems.
Firstly, enabling in-database AI-powered analytics to render data systems inherently AI-friendly proves to be intricate, particularly in accommodating modern AI applications like Large Language Models (LLMs) and generative AI.
Secondly, the vision of fully autonomous, intelligent data systems is still a distant reality, as existing data systems necessitate extensive manual oversight and maintenance.
Lastly, the fusion of AI and databases introduces new challenges in data privacy, which existing techniques often overlook.



\revision{We present \dbname, our initiative towards creating an AI-powered autonomous system that provides in-database AI-powered analytics to seamlessly support AI applications, and fully embraces AI techniques in each major system component to offer self-driving capabilities.}
We introduce methodologies and insights to address the above challenges, focusing on creating an in-database AI-friendly ecosystem, developing an intelligent self-driving data system, and providing privacy preservation and trusted AI$\times$DB.
For each 
objective, we delve into various new techniques employed in \dbname, including in-database dynamic model selection and slicing, fast-adaptive learned database components, in-database federated learning, etc.
We also outline future pathways for the development of \dbname, highlighting the open research challenges that must be addressed along this journey.


The structure of this paper is organized as follows. 
In Section~\ref{sec:background}, we present the relevant background to drive the discussion on next-generation data systems.
Section~\ref{sec:overview} defines key features and characteristics central to the next-generation data systems.
We detail our contributions to \dbname by introducing several key techniques in Section~\ref{sec:design}, Section~\ref{sec:auto}, and Section~\ref{sec:secure}. 
We discuss the development 
of \dbname 
in Section~\ref{sec:nextstep} 
and conclude in Section~\ref{sec:conclusion}.

\section{Background} \label{sec:background}

In this section, we describe the evolutionary trajectory of data systems, review the related work on AI$\times$DB, and discuss emerging AI-powered applications in healthcare and e-commerce, respectively.

\subsection{Evolutionary trajectory of data systems}

\begin{figure*}[t]
\centering
\includegraphics[width=0.85\linewidth]{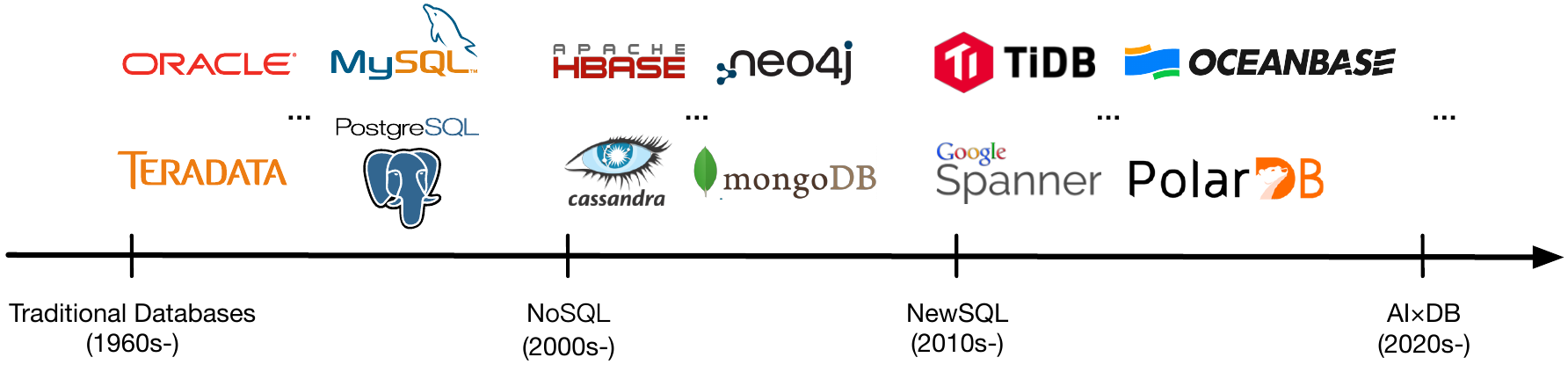}
\caption{\revision{Evolutionary trajectory of data systems}}
\label{fig:history}
\end{figure*}

The evolution of data systems can be segmented into several phases, each marked by innovations addressing the changing needs of data storage, management, and analysis.
As shown in Figure~\ref{fig:history}, these phases include the era of databases, the movement of NoSQL, and the emergence of NewSQL~\cite{DBLP:journals/sigmod/PavloA16}.

\textit{Phase 1: Traditional databases.} 
The origins of data systems can be traced back to the 1960s.
It was the introduction of relational databases in the 1970s that propelled data systems into the commercial spotlight.
Based on the relational model, relational databases excel at managing structured data, offering robust transaction support to ensure ACID (Atomicity, Consistency, Isolation, Durability) properties~\cite{DBLP:books/mk/GrayR93}, and providing a powerful query language, namely SQL (Structured Query Language). 
Due to these strengths, many of today's applications continue to rely on traditional relational databases, such as Oracle, MySQL, etc.
However, traditional databases typically adopt single-machine deployment, which restricts their scalability and flexibility.

\textit{Phase 2: NoSQL.}
As the volume, velocity, and variety of data began to expand, the limitations of traditional databases, particularly in terms of scalability, became apparent.
These problems sparked the NoSQL movement in the mid to late 2000s.
NoSQL systems were designed to offer scalable data processing, catering to the needs of big data and real-time online applications. 
They could handle unstructured and semi-structured data, supporting a variety of data models, e.g., key-value, document, wide-column, graph, etc. 
Despite their advantages, NoSQL systems often sacrifice transaction guarantees, leading to challenges in data consistency and integrity.

\textit{Phase 3: NewSQL.}
By the end of the 2000s, it was clear that many critical applications, such as e-banking and e-commerce, required both the scalability offered by NoSQL, as well as the ACID guarantees provided by relational databases.
To address this need, NewSQL systems were designed to distribute data across multiple machines for enhanced scalability, while offering robust transaction support to ensure ACID properties.
More recently, with the advancements in cloud computing, NewSQL has begun migrating to the cloud. 
This transition has given rise to many cloud-native databases that leverage the elasticity, resilience, and efficiency provided by the cloud.

\textit{Next-generation data systems.} 
The emergence of AI-powered applications across various domains is catalyzing the evolution of data systems towards a deeper fusion with AI.
Without this critical advancement, data systems will fail to offer automated and personalized data services demanded by modern AI models such as LLMs and generative AI.

\subsection{Related work on AI$\times$DB} \label{subsec:related_work}

The aspiration of integrating DB systems and AI technologies was first expressed some forty years ago~\cite{DBLP:conf/aaai/Brodie88}.
With the advancement in both fields over the years and the increasing demand for ``intelligent'' data systems offering advanced analytics, the fusion of both technologies become 
more imminent.
In 2016, the mutual benefits that deep learning and databases can offer each other have been
elaborated~\cite{sigmodRecord},
and more recently, various research works on AI$\times$DB have been proposed.

For one thing, databases can significantly enhance AI applications.
Databases have the potential to efficiently manage numerous AI models using their inherent user-friendly SQL interfaces~\cite{DBLP:journals/pvldb/XingCCLOP24}, and efficiently manage training/validation data for AI models~\cite{sigmodRecord}.
For another, AI can substantially optimize databases.
AI-driven techniques, such as knob tuning, can automatically refine database configurations to improve performance, which often surpasses traditional manual tuning done by database administrators (DBAs).
However, existing approaches on AI$\times$DB 
typically focus on optimizing within their specific domains,
and have not yet embraced a full fusion of AI and DB into a unified data system. 
We now delve deeper into the existing works to highlight the integration challenges and opportunities.




1) \textit{DB support for AI.}
Database techniques
under this category aim to
improve data quality for more accurate model training,
provide automatic model selection,
and facilitate model inference.
However, existing approaches are often built on top of the database rather than fully integrated within it, which limits their effectiveness and efficiency.
First, while various approaches, such as data cleaning, data integration, data discovery, data enrichment, and data lineage, are proposed to improve data quality, they are not directly supported by existing databases.
Implementing these techniques often requires complex combinations of multiple SQL commands, which can be user-unfriendly and inefficient.
Second, several systems~\cite{DBLP:journals/pvldb/XingCCLOP24,DBLP:conf/icde/AndersonC16} that are built on top of databases can automatically select suitable models for AI applications.
However, these systems often need to retrieve and transfer data from the database, which can be costly and time-consuming due to the increasing volume of data.
Third, recent proposals~\cite{ParkSBSIK22,inferdb} target to enhance model inference efficiency, however, they typically focus on traditional machine learning models, which may be ineffective for today's large-scale deep learning models.


2) \textit{AI support for DB.}
AI has shown potential for optimizing database systems. 
Previous works~\cite{DBLP:conf/sigmod/0006ZJWBLMP21,DBLP:conf/sigmod/ZhangW0T0022} introduce automatic knob tuning to improve system performance, but they provide limited insight into the possibilities of AI-powered database components, leaving broader opportunities for AI enhancements largely untapped.
Despite several efforts to create learned database components, such as learned query optimizers for more accurate query plans and learned indexes for faster key searches, these solutions rely heavily on offline training on static datasets.
This paradigm limits their effectiveness when dealing with dynamic workloads in real applications, rendering them impractical for real-world use.
Further, existing works often neglect the potential of combining AI with recent advancements in databases, such as the disaggregated architecture that separates computation from storage in cloud environments to improve scalability and elasticity.




\subsection{Emerging AI-powered applications} \label{sec:application}

AI-powered applications are increasingly becoming integral across various sectors, which aim to relieve the burden on end-users by offering personalized, automated AI-powered analytics and predictions. 
\revision{With \dbname, the construction of such AI applications can be simplified, and their efficiency can be improved.
Here we outline the transformative AI applications in healthcare and e-commerce
domains, and envision how they can be built with the unique functionalities of \dbname.}

\subsubsection{Intelligent healthcare}

Healthcare has continuously improved over the years, and with recent advancements in AI techniques, it is becoming smarter. 
We envision that the future of AI-powered healthcare promises exceptional intelligence, enabling more precise diagnoses, customized treatment plans, and proactive health management~\cite{DBLP:books/sp/17/LeeLNZZCOY17}.
AI-powered healthcare assistants will play a crucial role in assisting physicians in achieving more precise diagnoses. 
By leveraging AI models trained on extensive datasets, including genetic profiles, medical imaging, electronic medical records (EMRs), physicians' notes, etc., these assistants can reveal patterns not detectable by the human eye in areas such as neurology and orthopedics.
More importantly, these virtual assistants can offer 24/7 support to patients. 
They will handle inquiries about medication schedules, offer advice on managing chronic conditions, provide insights into physiological changes, and ensure adherence to treatment plans.

\revision{\dbname can simplify the establishment of intelligent healthcare applications.
Specifically, the digital diagnosis system, where doctors and physicians record EMR data and other relevant patient and treatment information, can be constructed based on \dbname, which provides a mature infrastructure for fast data access and processing.
Since all the required data is stored in \dbname, applications such as AI-powered healthcare assistants can directly perform the necessary and predictive analysis within \dbname using its in-database AI-powered analytic capabilities, which will be elaborated in Section~\ref{sec:overview}. In general, future intelligent healthcare applications can be efficiently managed by \dbname, eliminating the complexity of involving multiple dedicated systems for data management, model serving, etc.
}

\subsubsection{Intelligent e-commerce}

AI is revolutionizing e-commerce with automated fraud detection, personalized recommendations, and intelligent resource management tailored to workload demands.
By learning from transactional logs, AI models can uncover and predict anomalies to achieve accurate and real-time fraud detection, protecting businesses and customers from financial losses~\cite{DBLP:journals/pvldb/XiaoW00O23}.
Further, AI provides more efficient and enjoyable online shopping by offering personalized recommendations based on customer browsing history, purchase records, and preferences.
During high-traffic events such as Black Friday and Double Eleven (Singles' Day), AI-powered resource management is able to dynamically adjust the system to handle sudden spikes in demand, which guarantees the availability and efficiency of e-commerce systems.


\revision{We envision that \dbname can efficiently support intelligent e-commerce applications. 
For example, \dbname can support online purchases by maintaining complex relational tables of purchasing records and context attributes.
Based on such relational tables, we can conduct predictive tasks, such as link prediction to indicate potential purchases and fraud detection, directly within \dbname.
In particular, \dbname can automatically transform the relational purchasing table into a graph with user and item nodes, and then build deep learning models, such as GNNs or transformers, on this graph. 
This approach simplifies the process compared to traditional methods where graph construction and model development are performed outside the database, often resulting in higher costs and complexity.
Moreover, fine-tuning a pre-trained LLM based on the underlying domain-specific e-commerce data is also desirable and feasible within \dbname.
}

\section{Overview of \dbname} \label{sec:overview}

In this section, we describe the key principles of designing \dbname and present the conceptual and architectural overview of \dbname.

\subsection{Key design principles}

With the dynamic nature of workloads, data systems must inherently adapt to changes in data storage, data structures, system configurations, etc.
Traditionally, such dynamism is manually managed through actions like data insertion or update, index creation, and configuration tuning.
However, with the fusion of AI and DB, next-generation data systems anticipate a shift towards AI-driven autonomy, characterized by self-assembling and self-optimizing capabilities.
This shift entails that the entire system architecture can dynamically and autonomously adapt to varying demands under the guidance of AI.
For example, under heavy AI-enhanced analytical workloads, the system's embedded AI agent is expected to autonomously allocate more GPU resources and assemble appropriate AI models tailored to the analytical tasks.
In contrast, for standard transactional workloads that do not require AI-specific operations, the system might revert to a conventional, non-AI architecture.
These observations lead to our definitions of three essential properties for the design of next-generation AI-powered autonomous data systems.

\textit{Dynamicity.}
Dynamicity refers to the property that a data system can evolve autonomously in response to changing workloads and environmental conditions.
Dynamicity affects both the data managed by the system and the system architecture, including its components and operational behaviors.
This property involves a controlled system evolution from one state to another, ensuring a linear progression through these states.


\textit{Reliability.}
Reliability is defined as the ability of a data system to consistently meet performance and accuracy standards, even 
during adaptation and evolution.
That is, the data system is able to perform operations with optimal accuracy and consistency across all states.
However, reliability often requires a static or controlled environment to prevent unexpected state transitions that may compromise this property.

\textit{Scalability.}
Scalability refers to the system's ability to effectively increase capacity by either scaling up (enhancing existing infrastructure) or scaling out (adding more nodes/resources to the system).
This property 
guarantees that as the workload grows, the system can scale and/or reconfigure 
to maintain or improve performance, while 
achieving strategic resource allocation and utilization.



Data systems are dynamic in nature as the data distributions and query patterns evolve over time, and therefore, the systems have to adapt for the dynamicity while guaranteeing reliability.
We are therefore using the Filter-and-Refine Principle (FRP) to guide the design of \dbname.

\textbf{FRP conjecture.} For a dynamic and real-time data system, computational models or algorithms must be designed based on the Filter-and-Refine Principle, so that the system only considers the data subspace or strategies that potentially yield the near-optimal performance.


\noindent Due to the unpredictable nature of real-world environments and the inherent uncertainties in prediction accuracy and latency, the data systems must adapt and provide acceptable performance.
The system optimization workflow often incurs several iterations.
%
%
For a dynamic system with huge amounts of data, FRP is the basis for adaptivity, efficiency, and scalability.
%
The filter-and-refine strategy is typically used to speed up computationally intensive tasks so that they can be completed within the pre-defined latency and/or resource constraints.
In some tasks, such as a join, we have to remove data that are surely not contributing to the answer quickly, and then perform computationally intensive steps on a much smaller subset of the original data.
Likewise, we eliminate query processing plans that are obviously suboptimal quickly, and only look for a near-optimal plan. For iterative computations, we should take actions that move to the target within a smaller search space.

\subsection{Key features} \label{sec:challenges}

Next-generation data systems are confronted with three critical features that must be achieved to efficiently and effectively support AI-powered applications.


\subsubsection{In-database AI-powered analytics}

AI-powered analytics are fundamental in constructing emerging AI applications, as discussed in Section~\ref{sec:application}.
Unlike traditional data analytics that typically offers a limited statistical view based on filtering or aggregation operations~\cite{singa,singaeasy,jia2023robust}, AI-powered analytics is more dynamic to capture and discover the complexity and intricacies of underlying data patterns~\cite{gray1997data}.
However, existing approaches to AI-powered analytics~\cite{rafiki,clipper} fall short in efficiency and effectiveness, primarily due to the inherent limitations of using separate data systems and AI models.
Critical operations required by AI models, such as data preparation, model training, and making inferences, necessitate extensive interactions with data systems to retrieve the corresponding data. 
With the continuous emergence of new AI models, such as Deep Neural Networks (DNNs)~\cite{RESNET,NASNET} and LLMs, such interactions become increasingly costly due to the vast amounts of data these models require.

To address this, we assert that in-database, AI-powered analytics 
is
crucial and should be a key feature of \dbname. 
Consider a user named Tom browsing an e-commerce app.
For a personalized recommendation, we aim to enable \dbname to directly answer complex analytical queries such as ``What are the top 10 recommended items for Tom?'' by automatically selecting the appropriate AI model and delivering results without costly data transformation.
Such capability would significantly enhance the user-friendliness, effectiveness, and efficiency of AI-powered analytics.

\subsubsection{Intelligent self-driving data system}

Enhancing database systems with AI techniques is becoming increasingly popular. 
For example, automatic knob tuning can optimize database configurations by selecting the best system parameters.
Similarly, data structures like indexing can be improved with learning methods that adapt to data distribution patterns.
However, existing approaches primarily focus on using AI for knob tuning or optimizing specific system components based on specific logs or patterns.
As a result, their designs often appear 
heuristic and static, failing to provide thorough optimization that fully interacts with and enhances the system. 
It still lacks a holistic approach where AI is the first citizen to fully exploit AI's potential in data systems.

We envision that in-database AI-powered analytics will catalyze a comprehensive evolution of data system architecture and techniques, leading to a fully autonomous data system that can dynamically adapt itself with little or even without human intervention.
This transformation requires AI optimizations to make the data system scalable, adaptive, and intelligent.


\subsubsection{Privacy-persevering and trusted \aixdb}

As the need for analytics on sensitive data and data collaboration intensifies~\cite{DBLP:journals/pacmmod/WangWC0O23}, ensuring data privacy and trustworthiness has become indispensable.
Existing privacy-persevering techniques typically focus on AI or databases separately, overlooking the specific challenges presented by AI$\times$DB. 
On the one hand, while privacy-enhancing techniques work in the AI domain, such as federated learning~\cite{DBLP:journals/pvldb/WuX0DLOX023,DBLP:journals/pvldb/WuCXCO20} that enables multiple parties to collaboratively train models and make inferences, these approaches are not directly applicable to in-database AI models. 
On the other hand, traditional database security methods~\cite{DBLP:journals/pvldb/WangDLXZCCOR18}, such as data masking and auditing, are insufficient to protect the data during the interaction with AI components. 

We envision that new privacy issues will emerge during the development of \dbname, requiring the data system to natively support data and AI model management in a privacy-preserving and trusted manner.


\subsection{Conceptual framework of \dbname} \label{sec:framework}

\begin{figure*}[t]
\centering
\includegraphics[width=0.86\linewidth]{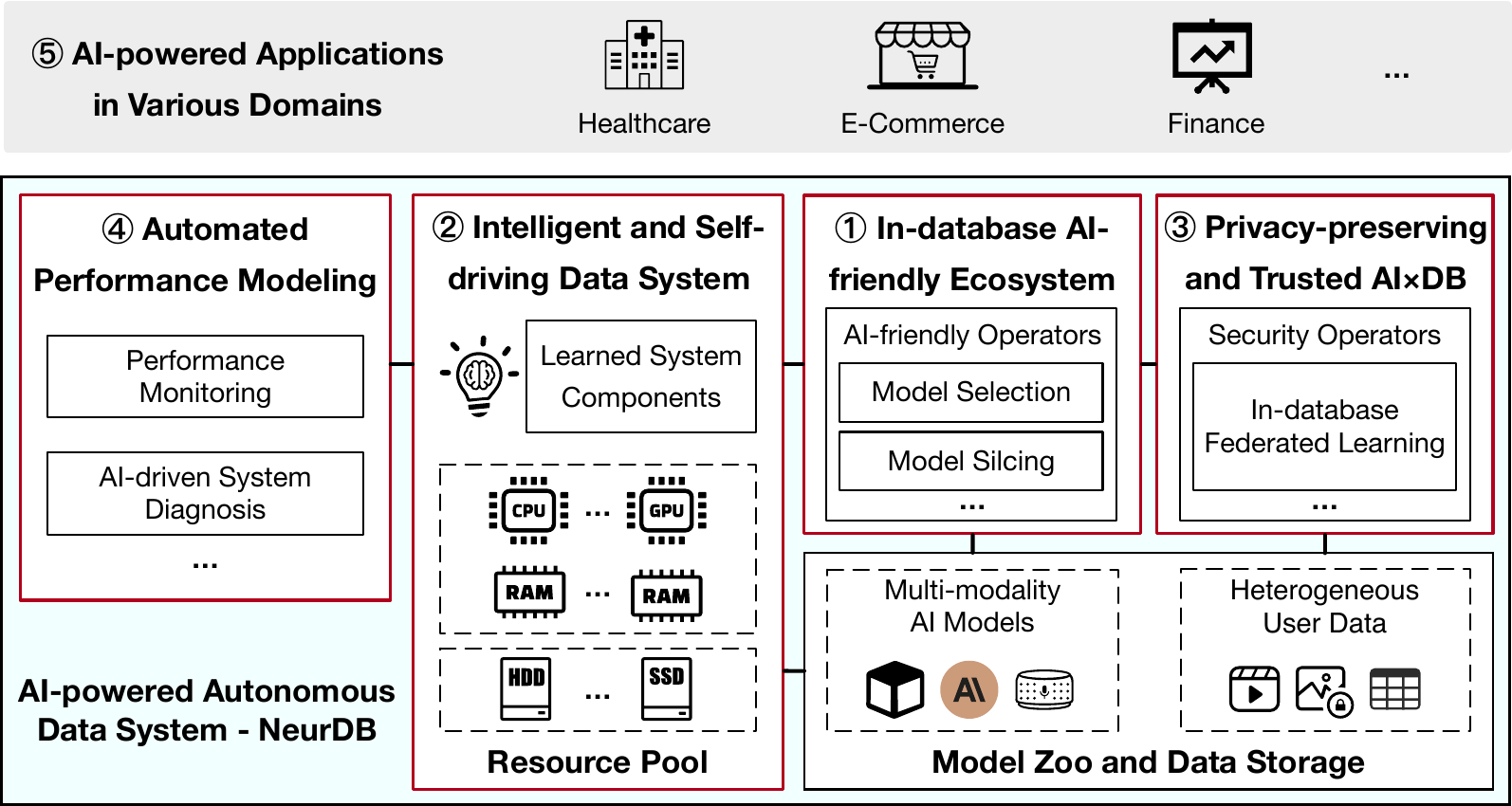}
\caption{\revision{Conceptual framework of \dbname}}
\label{fig:overview}
\end{figure*}

To provide these key features, we establish the conceptual framework of \dbname in Figure~\ref{fig:overview}, which underpins its overall design and functionality.
\revision{\dbname consists of four key components: \ding{172} in-database AI-friendly ecosystem, \ding{173} intelligent self-driving data system, \ding{174} privacy-preserving and trusted AI$\times$DB, and \ding{175} automated performance modeling.
These components correspond with and interact with each other to effectively and efficiently support \ding{176} AI-powered applications in various domains.}

\subsubsection{In-database AI-friendly ecosystem}

We advance beyond existing works by ensuring that \dbname can inherently support AI applications, thereby improving model accuracy, efficiency, and usability. 
We focus on two aspects: 
a) Developing easy-to-use, AI-friendly functionalities in \dbname. 
We enhance database functionalities to facilitate efficient and user-friendly AI model management, i.e., allowing users to conveniently use AI models and retrieve the required data by models; 
b) Investigating efficient in-database model training and inference techniques. 
First, we support in-database neural architecture search to minimize data transfer overhead, ensuring more efficient model training. 
Second, we explore in-database model slicing techniques to serve as a query optimizer for AI models, to enhance model inference efficiency. 
Third, we propose to design component-based modeling, such as modular neural networks, enabling a mix-and-match approach for improved efficiency and accuracy of AI models. 

\subsubsection{Intelligent self-driving data system}

We fundamentally redesign the database architecture and propose AI-driven database optimizations tailored for AI-powered hybrid transactional and analytical processing (AI+HTAP). 
To this end, we focus on investigating scalable intelligent database design and advanced AI-driven database optimization techniques. 
First, we aim to identify the requirements for making AI an integral part of \dbname, and propose a new AI-powered data system architecture.
Second, we incorporate AI components with the recent evolution of databases, such as cloud-native disaggregation, new hardware, etc., to make \dbname scalable and self-driving. 
Third, we ensure high-performance AI+HTAP by designing specific data structures to provide efficient and effective AI functionalities, such as in-database Retrieval-Augmented Generation (RAG), etc. 

\subsubsection{Privacy-preserving and trusted AI$\times$DB}

We develop new techniques to address the issues that arise with \aixdb for enhanced data protection.
Our focus is on two main areas: 
a) Privacy-preserving AI$\times$DB. We propose incorporating secure operators (e.g., secure multiparty computation protocols) into databases to facilitate in-database federated learning for strong privacy protection with high efficiency. 
Moreover, it is crucial to ensure that AI models are not vulnerable to various inference attacks. 
Hence, we aim to design mechanisms that offer in-database differential privacy to protect the data utilized for training the AI models. 
b) Trusted AI$\times$DB. We propose integrating the verifiable data structure into the native database engine to ensure the integrity of data accessed by in-database AI models. 
Further, we propose employing zero-knowledge proof (ZKP) techniques to verify that the computational results derived from the data are well-formed and trustworthy.

\subsubsection{Automated performance modeling for AI$\times$DB}

New automated validation approaches are required to ensure the correct, secure, and performant functioning of \dbname.
However, existing methods face limitations: they assume the data system will produce deterministic results, rely on self-contained test cases, and expect a stable system state.
This assumption does not apply to systems like \dbname, which is AI-driven, adapts to its workloads, and natively supports various AI workloads.
To tackle these issues, we propose to design a new adaptive and automated validation system that can automatically generate test inputs for \dbname to validate the system performance, security and fairness, and the effectiveness and efficiency of the supported AI applications.
In addition, to gain insights into usability and deployment challenges, we plan to explore real-world deployment opportunities to validate the practicality of our proposed systems and techniques. 
We plan to utilize \dbname to support actual AI application development, such as AI healthcare analytics developed by clinicians for their data, and implement \dbname in industry settings, by working with cloud service providers and e-commerce vendors.


\subsubsection{AI-powered applications in various domains}

We plan to develop a set of application-oriented toolkits to provide built-in AI as a service for non-experts in AI and computer science, especially for users in the healthcare and e-commerce domains. 
Healthcare, with its complex data and growing need for accurate and personalized predictions, and e-commerce, with its vast consumer data and dynamic user needs and market trends, represent two critical domains where the integration of \aixdb can yield transformative results for enhancing user experience and improving operational efficiency~\cite{zheng2020tracer,DBLP:conf/icde/CaiZOWY22}. 
Current practices in these domains treat data management and AI as separate components, requiring domain knowledge in both areas~\cite{DBLP:conf/sigmod/Zheng0HNOG21}. 
Such a segregated solution limits the potential of AI applications and creates barriers to innovation and vast application.
We aim to build our next-generation \aixdb system to support more personalized, adaptive, and scalable healthcare and e-commerce applications: 
a) Declarative interface with minimal domain knowledge requirement. \dbname aims to introduce a user-friendly, declarative interface, akin to LLMs, which simplifies the interaction with AI models. 
b) Usability and scalability for large-scale data.
Through \dbname's unique AI-powered analytics, we aim to improve applications like Foodlg, which we designed to monitor and balance patients' diets, to support the entire treatment process autonomously with \dbname's integration.


\section{In-database AI-powered analytics} \label{sec:design}


In this section, we elaborate on the architecture of the in-database AI-friendly ecosystem and detail the key techniques that we have proposed to support in-database AI-powered analytics for \dbname.

\subsection{AI-friendly ecosystem overview} \label{sec:overview_analytics}

\begin{figure}[t]
\centering
\includegraphics[width=0.99\columnwidth]{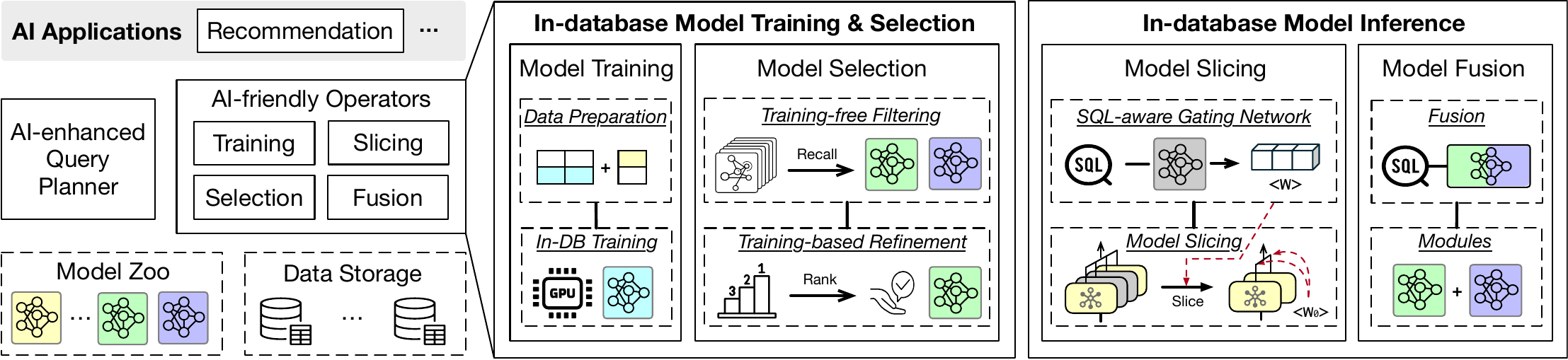}
\caption{Overview of in-database AI-friendly ecosystem}
\label{fig:overview_analytics}
\end{figure}

We establish the AI-friendly ecosystem of \dbname following the architecture illustrated in Figure~\ref{fig:overview_analytics}.
At the core of this ecosystem is an AI-enhanced planner, tasked with various automatically activating AI-friendly operators to process complex analytical queries.
These operators integrate the entire workflow of AI applications within the data system, which ensures \dbname is inherently AI-friendly.
In \dbname, we introduce two main categories of AI-friendly operators: model training \& selection operators and model inference operators.
1) Model training \& selection operators are designed to ensure task-specific model effectiveness without costly data transformation.
In particular, the model training operator invokes the training process for specified tasks within the database, while the model selection operator aims to obtain the best-performing model with the optimal network topology for each task.
2) Model inference operators are employed to perform accurate and efficient inference. 
Specifically, the model slicing operator adjusts the AI model to better suit the given analytical query, and the model fusion operator enables modular neural networks that combine multiple models with a mix-and-match technique.


Consider the recommendation task in e-commerce as an illustrative example.
\dbname efficiently and effectively supports such tasks with its AI-friendly functionalities.
For personalized item recommendations,  the planner of \dbname can automatically generate an execution plan that might include a model selection operator followed by a model slicing operator.
In particular, the model selection operator identifies an appropriate model architecture for the task from the model zoo, which is then used to perform the inference.
However, the selected model, often generalized to capture common patterns and trends applicable to all users, may not achieve the highest accuracy for personalized recommendations due to its lack of focus on individual user characteristics, e.g., age and income.
To address this, we employ the model slicing operator to automatically refine the model, ensuring it better aligns with specific user characteristics for more accurate recommendations.
Moreover, \dbname might perform a model fusion operator, which integrates models from related tasks, such as a model for the click-through rate (CTR) prediction, to boost recommendation accuracy.

Integrating AI functionalities into databases requires more than simply embedding AI models within them~\cite{mlcask,pipeEval,compoundai,llmbpm}.
Besides supporting AI-friendly operators, data systems also need to efficiently track model evolution, manage pipelines, and interact with other internal/external tools, e.g., dynamic model versioning, automated pipeline generation, cross-model interaction, etc.
Such system level interactions beyond monolithic AI models have been implemented 
in some of our solutions such as MLCask~\cite{mlcask}, which functions like a Git-like system for managing the entire machine learning lifecycle, and IP8Value~\cite{ip8value}, which is
an application specific data system that leverages AI and LLM for intellectual property management.


We now detail the key techniques that \dbname employs to build the AI-friendly ecosystem.

\subsection{Dynamic model selection}
\label{subsec:selection}

Given the costly data transformation required by existing systems, we are therefore motivated to support in-database dynamic model selection.
However, providing efficient and effective in-database model selection is not trivial.
Existing solutions are typically categorized into two types: training-based and training-free.
Training-based algorithms~\cite{NASNET,Cerebro} can effectively measure the model performance but typically require training and evaluating hundreds to thousands of candidate models, which are computationally prohibitive in both time and resources~\cite{googlemodelsearch}.
In contrast, training-free model selection algorithms~\cite{zero-cost,hnas,OMNI} estimate model performance by computing certain statistics of candidate models without training, namely Training-Free Model Evaluation Metrics (TFMEMs).
While leveraging TFMEMs allows for the quick evaluation of numerous models, these approaches often compromise accuracy, potentially limiting the discovery of the most effective models compared to training-based algorithms that offer more precise evaluations.




Unlike existing works, we aim to develop a model selection algorithm for \dbname that harnesses the advantages of both training-free and training-based paradigms.
Our idea is to employ training-free algorithms to filter down the candidate model set, and then use training-based algorithms for refinement, i.e., obtain the most effective model.
Recognizing the inherent trade-off between training-free and training-based paradigms, careful coordination of these two phases is essential.
To address this, we incorporate Response-time Service-level Objectives (SLOs)~\cite{mark,PIQL}, which are strict response-time thresholds predefined by users.
Failing to meet the SLOs can lead to reduced query service quality or even financial losses~\cite{mark}.
Based on SLOs, we allocate specific time budgets for both the training-free and training-based phases to properly balance efficiency and effectiveness.



\begin{figure*}[t]
\centering
\includegraphics[width=0.92\columnwidth]{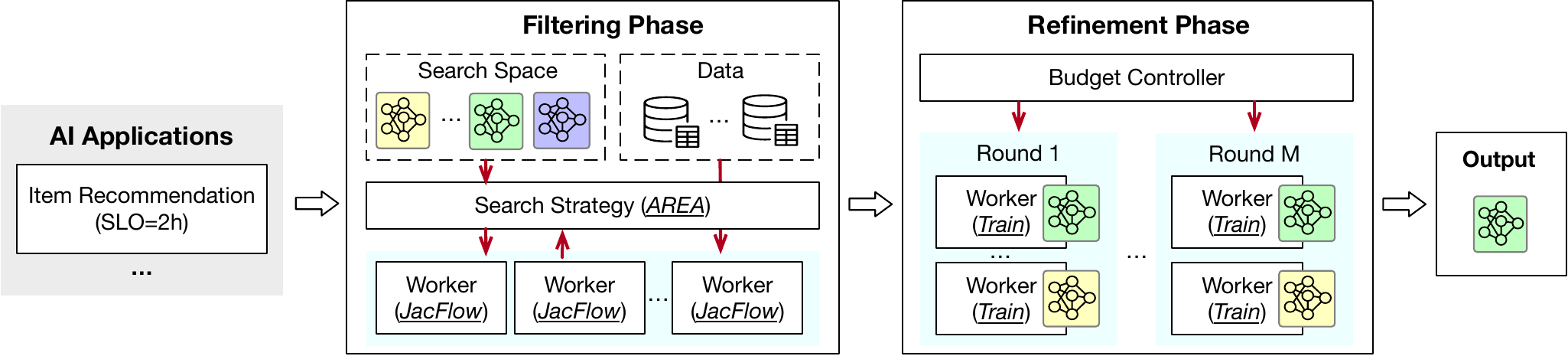}
\caption{SLO-aware dynamic model selection}
\label{fig:ms_alg}
\end{figure*}






We propose an SLO-aware model selection algorithm, which consists of a training-free filtering phase and a training-based refinement phase as shown in Figure~\ref{fig:overview_analytics}.
First, in the training-free filtering phase, we efficiently explore a candidate set of promising models with higher scores.
Specifically, we utilize an enhanced TFMEM named JacFlow~\cite{tabnas} to calculate a score for each model under consideration. 
JacFlow integrates expressivity-based~\cite{SYNFLOW} and trainability-based~\cite{naswot} TFMEMs to improve model ranking accuracy.
Second, in the training-based refinement phase, we determine the optimal model from the candidate set through multiple rounds of examination.
Each round involves training the candidate models for a specified number of epochs, progressively narrowing the candidate set for the next round's evaluation.
Further, to align this process with SLO, we have developed an SLO-aware function that manages three critical parameters: the number of models to be scored in the filter phase ($N$), the candidate set size ($K$), and the training epochs used in the refinement phase ($U$).
By so doing, we ensure the selection of a higher-performing model while meeting the predefined response-time threshold. 
For further details on the SLO-aware model selection algorithm, please refer to~\cite{DBLP:journals/pvldb/XingCCLOP24}.



As shown in Figure~\ref{fig:ms_alg}, we present a running example of the model selection process with an SLO of two hours for the item recommendation task discussed in Section~\ref{sec:overview_analytics}.
During the filtering phase, we obtain a candidate set of promising models.
To enhance efficiency, we deploy multiple workers to score the models using JacFlow in parallel.
Given the potential of numerous models, scoring each model is infeasible and costly.
Thus, we implement the asynchronous regularized evolution algorithm (AREA)~\cite{HIEAS} to search for the most promising models for scoring and optimally distribute these models among workers. 
The filter phase terminates once $N$ models are scored, from which the top $K$ scored models are selected to form the candidate set.
In the subsequent refinement phase, we train and evaluate the models in the candidate set with multiple rounds.
In each round, each model in the candidate set is trained for $U$ epochs and then assessed to retain only the top $1/\eta$ performing models, with $\eta$ being a predefined factor regarding the epoch count.
We initiate the first round with $U=U_{init}$, and in subsequent rounds, $U$ is adjusted to $U' \cdot \eta$, where $U'$ denotes the epoch count used in the previous round. 
This incremental training approach allows each model to receive a progressively larger training budget in successive rounds, facilitating a more precise evaluation.
In this way, we finally obtain the best-performing model.

\subsection{Dynamic model slicing}


Model slicing~\cite{DBLP:journals/pvldb/Cai0OG19} is a critical technique for adjusting AI models to meet the specific requirements of various analytical tasks. 
It customizes the architecture of deep learning models based on analytical queries, thus improving the model inference accuracy.
As data volumes grow, analytical queries often focus on specific subsets of the data rather than the entire database.
Traditionally, a single general model is used to support inference for different analytical queries~\cite{he2017neural,khamis2020learning}.
However, such a model, optimized to capture common patterns across the entire database, may not deliver as accurate inferences as a dedicated model trained on a specific subset of interest.
For example, when examining patient data, analysts prefer to study patterns or trends among individuals diagnosed with a particular disease. 
A model exclusively trained from the data of diabetic patients might provide more precise predictions than the general model for all patients regarding their readmission rates.
Given the vast number of possible analytical queries, training a dedicated model for each one is also computationally impractical.
To this end, we consider leveraging query conditions to perform model slicing.
In particular, for each analytical query, we analyze and encode the query conditions, which indicate the specific data required for the query, and then use this information to adjust the model architecture accordingly.


\begin{figure}[t]
\centering
\includegraphics[width=0.95\columnwidth]{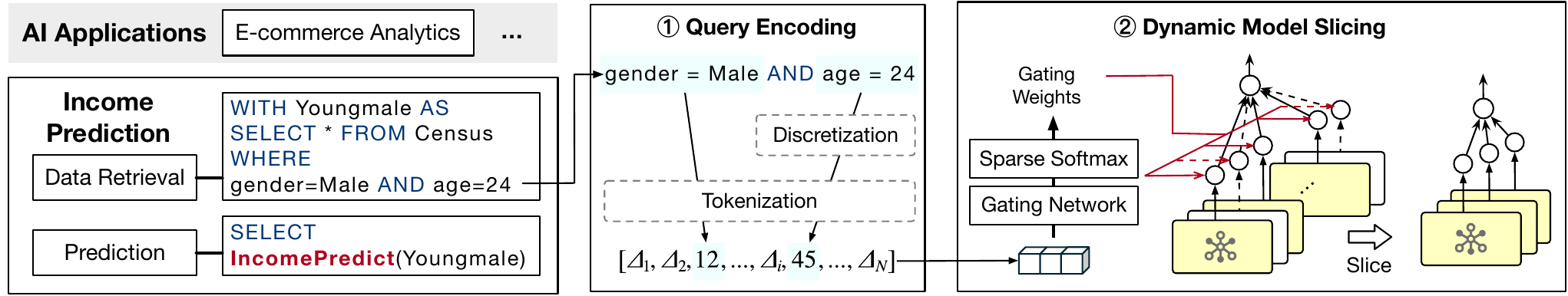}
\caption{SQL-aware dynamic model slicing}
\label{fig:slice}
\end{figure}

We proposed LEADS~\cite{zeng2024slicing}, a new SQL-aware dynamic model slicing technique that enables the customization of the predictive model for a subset of data retrieved by analytical queries.
The core of LEADS has three parts: (1) the construction of the general model, (2) the encoding of the analytical query, and (3) the slicing via SQL-aware gating network. 
The goal of the first step is to improve the capability of the predictive base model. We scale it horizontally following the Mixture-of-Expert (MoE) technique to form a general model. The general model is composed of multiple replicas of the base model. These replicas, termed experts, are trained to specialize in different problem subspaces for more effective predictive modeling. Given the predictive base model as $\mathcal{F}(\mathbf{x})$, the general model is represented as  $\mathcal{F} = [\mathcal{F}_1,\mathcal{F}_2,\cdots, \mathcal{F}_K ]$, where K is the number of experts.
In the second step, we model the filter condition in the analytical query. Considering the potential combination of filter conditions, for simplicity, we focus on individual analytical queries, referred to as primitive queries.
In the primitive query, each attribute should be associated with zero or one predicate, with predicates across attributes conjoined using the logical operator $\land$ (\texttt{AND}). The $i$-th attribute value of predicates is denoted by $q_i$.
Particularly, a padding value is used if there are no predicates in attributes.
Given 
$N$ attributes in the dataset, we can obtain a fixed-dimensional embedding of the analytical query, $\textbf{q}$, by concatenating the attribute values of the predicates: $\textbf{q} = [q_1, q_2,\cdots, q_N]$.
Further, 
a SQL-aware gating network $\mathcal{G}$, a combination of a two-layer neural network and a sparse softmax layer,
is employed.  
For a specific analytical query $\textbf{q}$, the SQL-aware gating network dynamically generates gating weights for each expert in the general model, denoted as $\textbf{w} =\mathcal{G}(\textbf{q}) = [w_1, w_2, \cdots, w_K]$.
$w_i = 0$ indicates that the corresponding $i$-th expert is not required in the current predictive modeling. Thus, we only activate a small fraction of experts $\mathcal{F}_i$ for the model inference. 

Figure~\ref{fig:slice} shows a practical application of LEADS in business analytics. 
Consider an analytical query that predicts the income of 24-year-old men. 
The query's execution can be divided into two phases: data retrieval and prediction. 
During data retrieval, filter conditions are applied and undergo query encoding. 
As depicted in Figure~\ref{fig:slice}, the categorical value ``gender = Male'' is directly mapped to a token.  
Meanwhile, the numerical value ``age = 24'' is discretized into a range before tokenization. 
If a column lacks a filter condition, the corresponding position is padded with a default token. 
This process yields a fixed-sized query token capturing metadata about incoming predictive data.
Subsequently, we customize the general model via model slicing. 
The query embedding is fed into a gated network. 
Leveraging a well-designed sparse softmax, recalibrated gating weights are obtained, where low scores are set to zero. 
Based on these gating weights, the experts within the scaled-up general model are activated selectively. 
Experts with zero scores are disregarded, and the remaining activated models constitute the slice model. 
This sliced model is then applied to the subsequent predictions, facilitating insightful analytics tasks.

\subsection{Dynamic model construction with plugins}

Component-based modeling is a neural network technique designed to manage AI models by dividing them into several modules.
This technique enables independent development of each module, which can be easily added, removed, or replaced in a plug-and-play manner. 
Such flexibility and extensibility facilitate easier model implementation, upgrades, and maintenance.
We have utilized component-based modeling in the healthcare domain and demonstrated its benefits for data analytics in our previous work~\cite{DBLP:journals/pvldb/ZhengCCHZO22,DBLP:journals/pvldb/WangOYZZ14,DBLP:conf/mm/WangCDGOTW15}.
There are also several related works to support component-based modeling in various other domains.
However, as AI models become more advanced and complex, existing approaches face two key limitations.
First, given today's diverse AI models, there may exist a large number of modules for various purposes.
Consequently, efficiently managing these modules while ensuring flexible and extensible model construction at runtime is challenging.
Second, establishing these modules often requires specific domain knowledge such as functions, data flows, or analytical pipelines, and therefore, this process is rather complex for end users with limited domain knowledge.


Unlike existing works, 
we aim to design a data-driven and database-native component-based modeling technique to make \dbname generalizable for various AI-powered analytics.
Our basic idea is to leverage the fundamental features offered by databases, such as efficient query processing and data storage, to embed and integrate the entire component-based modeling process within the database.
However, solving the problem of efficient module management and user-friendly module designs is not trivial.
We first create tailored storage for module management, which includes an index specifically designed for a large number of modules.
Building on this, we then develop a model fusion operator for efficiently retrieving the required modules during the execution of analytical queries.
We extend the query executor to support this model fusion operator.
The fusion operator is executed following the widely used volcano execution model, which first processes independent modules in parallel, and then applies multimodal fusion to capture cross-module interactions.
To improve user-friendliness, we enhance the query optimizer to guide the model fusion.
In particular, the model fusion operator can be automatically triggered upon receiving a complex analytical query.
This enhanced query optimizer generates a proper execution plan including the appropriate modules needed for this query.
Further, we plan to employ the query \texttt{analyze} and \texttt{explain} syntax to enhance the interpretability of the model fusion process.





\section{AI-powered autonomous data system} \label{sec:auto}

In this section, we present the architecture of autonomous \dbname, and detail the key components that we have developed to transform \dbname into an intelligent self-driving data system.


\subsection{Autonomous architecture overview} \label{subsec:auto_overview}


The ideation
of making databases autonomous and adaptive was made in various forms some forty years ago~\cite{DBLP:conf/sigmod/HammerC76,DBLP:conf/sigmod/GraefeD87}, and it has been revisited over times~\cite{DBLP:conf/sigmod/0006ZJWBLMP21,DBLP:conf/icac/GuptaMD08}.
A fully autonomous data system typically refers to a system that can dynamically adapt itself without human intervention.
Such a system is expected to determine appropriate actions, execute them, and learn from the outcomes to achieve dynamicity.
These actions can include performance optimization, system repairs like replacing a failed node, and other self-maintenance tasks.
Although significant progress has been made toward achieving a fully autonomous data system, existing systems still require some level of human intervention.
For example, system repairs can be triggered automatically but often require human validation to avoid data inconsistency.
However, with the rapid advancement of AI techniques and new database architectures, it is 
about time
to design
a fully autonomous data system by leveraging these new developments and foundations.



\begin{figure*}[t]
\centering 
\includegraphics[width=0.89\columnwidth]{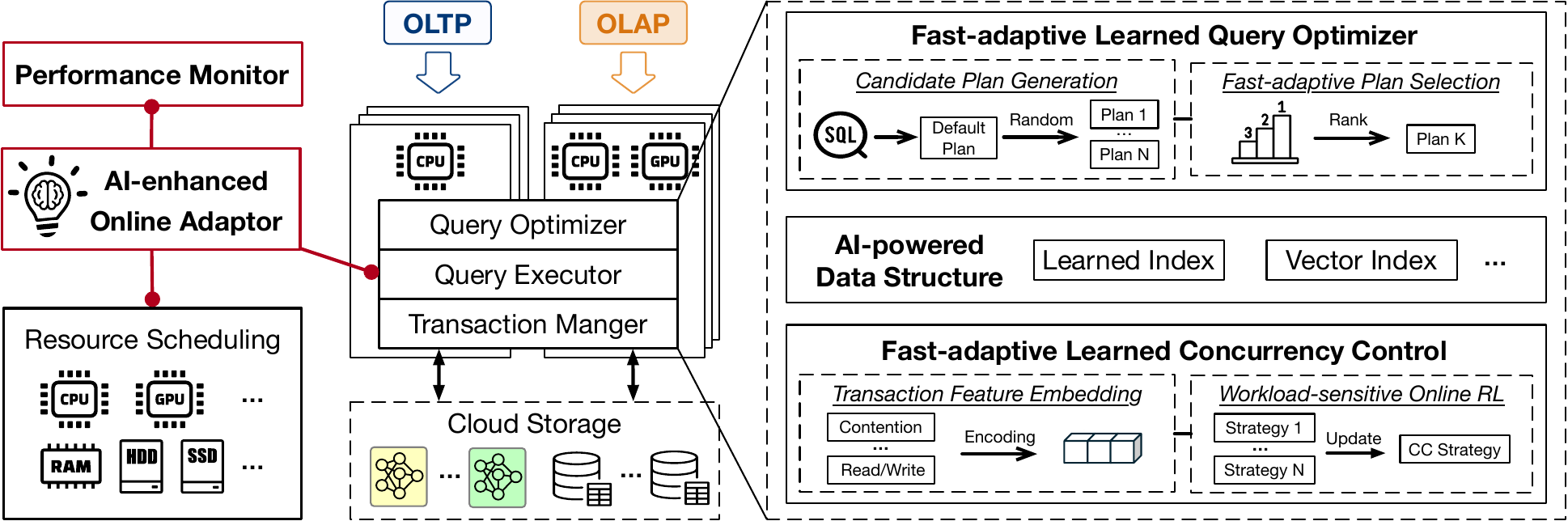} 
\caption{Autonomous architecture of \dbname} 
\label{Fig:overview_system} 
\end{figure*}

We propose a new autonomous data system architecture for \dbname, shown in Figure~\ref{Fig:overview_system}.
This architecture is disaggregated into two distinct layers: the computation layer and storage layer, enabling each layer to scale independently~\cite{DBLP:journals/pacmmod/ZhaoPCDLO23}.
Such disaggregation facilitates the system's efficiency, elasticity, resilience, and scalability.
Each layer consists of multiple nodes: in the computation layer, nodes process transactions and deliver results to users, while in the storage layer, nodes are dedicated to data storage.
\dbname includes essential operational components in the standard database design of the computation layer, such as the query optimizer, query executor, transaction manager, etc.
These components are 
facilitated by
AI to boost not only performance but also functionalities. 
For instance, the query optimizer in \dbname is now tasked with co-optimizing plans for both AI operators and traditional data operators, e.g., JOIN, incorporating insights from the AI-powered planner discussed in Section~\ref{sec:overview_analytics}.
To accommodate the diverse and dynamic workloads in real-world applications, we introduce an AI-enhanced adaptor to enable \dbname as a self-assembling and self-optimizing data system.
With the flexibility of the disaggregated architecture, there are more optimization spaces and strategies that can be explored by the adaptor.
In the computation layer, the adaptor intelligently assembles system components, such as selecting appropriate indexes and adjusting the concurrency control strategy, and allocates resources dynamically in response to specific workload demands.
In the storage layer, the data layout is automatically optimized based on workload requirements, with data co-location adjustments made to ensure fast access.




We now illustrate the autonomy of \dbname by discussing its efficient support of dynamic AI+HTAP workloads.
As shown in Figure~\ref{Fig:overview_system}, \dbname creates specialized nodes tailored to the demands of various workloads in the computation layer.
For online analytical processing (OLAP) workloads, nodes are provisioned with both CPU and GPU resources to handle the computational intensity.
In particular, for AI-powered analytics, \dbname configures nodes that combine extensive CPU and GPU resources, coupled with an AI-optimized query optimizer.
It employs high-dimensional learned indexes, suitable for RAG, and adopts optimistic concurrency control to manage the anticipated low-contention OLAP workloads.
In contrast, for online transactional processing (OLTP) workloads, we would allocate nodes equipped primarily with CPU resources to optimize resource utilization, as transactional tasks can be efficiently processed without the need for GPU support.
\dbname opts for B$^+$-tree indexes and employs lock-based concurrency control to handle potential high-contention reads and writes in OLTP workloads.
Regarding the storage layer, the adaptor can automatically organize a replica of data into a column format for OLAP workloads, which is more suitable for OLAP query patterns, while keeping other replicas in row format.



Developing such an efficient and effective autonomous data system is not trivial, particularly given the dynamic nature of workloads.
The major challenge is that as data in the database evolves over time, static AI models can become inefficient in providing reliable results.
For example, while learning from specific instances of datasets and query patterns can potentially obtain improved query plans, such improvements may be short-lived due to workload shifts and data updates. 
\revision{We therefore aim to design dynamic and online AI models that can be updated and adapted to data variations to maintain accuracy, thereby facilitating continuous optimization of system components for better performance.
Given that the Filter-and-Refine paradigm (FRP) is used as a common-known strategy in indexing~\cite{DBLP:journals/tkde/BertinoO99,DBLP:journals/tods/JagadishOTYZ05,DBLP:journals/tkde/TheodoridisSS00}, query processing~\cite{DBLP:books/mg/SKS20,DBLP:conf/ideas/OoiPWWY02}, etc., our core idea generally follows the FRP and employs online reinforcement learning (RL) to facilitate dynamicity.}
RL naturally aligns with FRP, as it involves filtering through a range of possible actions and then refining the most effective strategies based on feedback.
In particular, FRP's filtering stage is akin to RL's exploration phase, where the RL agent tries different actions to understand their effects on the environment, gathering information to guide future decisions. 
Further, 
the exploitation phase of RL shares the same objectives as the refinement step of the FRP, where the RL agent refines its strategy by focusing on actions that yield higher rewards. 
During this phase, the RL agent also optimizes its strategy to maximize long-term gains by learning from past experiences.
With these design principles in mind, we now elaborate on our key techniques for the query optimizer and concurrency control to realize seamless online adaptation in \dbname.


\subsection{Fast-adaptive learned query optimizer}
\label{subsec:optimizer}



The query optimizer is a crucial component of database systems, tasked with selecting the most efficient execution plans for given SQL queries. 
Traditional query optimizers are cost-based and typically consist of three main components: plan enumerator, cardinality estimator, and cost model.
Due to the difficulty of cracking the cardinality estimator and cost model, there is a growing interest in harnessing machine learning to enhance existing query optimizers~\cite{zhu2023lero,marcus2021bao,yang2022balsa,doshi2023kepler,marcus2019neo}.
Although those learned query optimizers demonstrate advantages over traditional ones in certain applications, they are not yet practical for real-world database systems.
First, machine learning models might overfit to the training data, resulting in poor adaptability to new or dynamic query workloads. 
Second, existing work adopts periodic retraining to incorporate new data, as a result, models can forget previously learned knowledge when updated with new information.

We therefore propose an effective and fast-adaptive learned query optimizer.
For the effectiveness, the optimizer must effectively find a near-optimal execution plan with lower query response time for a single query.
For fast adaptivity, the optimizer must quickly learn from limited experience related to unseen queries and can remember the learned knowledge without suffering from catastrophic forgetfulness.
To achieve this, we divide the proposed learned query optimizer into two stages: efficient candidate execution plans generation, and fast-adaptive plan selection via online reinforcement learning, as shown in Figure~\ref{Fig:overview_system}.

During the plan generation stage, we generate a set of potential execution plans, while keeping the set size small and ensuring the inclusion of plans that are close to optimal.
To this end, we obtain candidate execution plans by mutating the optimizer's cardinality estimates rather than relying on a pre-fixed set of hints as in existing works~\cite{marcus2021bao}.
In particular, we first employ the default optimizer to obtain a base execution plan for a given query.
We next construct a set of varied cardinalities by applying multiplicative factors to the original cardinalities provided in the base plan.
After that, we uniformly sample from these mutated cardinalities and feed them together with the original query into the default optimizer again to generate a new execution plan.
This process is repeated $N$ times to produce $N$ candidate execution plans.
Each plan may differ in execution strategies, such as join order, join algorithm, index usage, etc., due to the varied cardinality estimates.

In the plan selection stage, we utilize online reinforcement learning to select the near-optimal execution plan, while achieving fast-adaptive to unseen queries and remembering the learned knowledge without suffering from catastrophic forgetting.
Specifically, we model the execution of queries as a sequence of events and use a sequence model, i.e., decision transformer\cite{xu2023hyperdecision, chen2021decision}, to select the near-optimal execution plan.
For a given query and its corresponding candidate plans, we predict which plan may yield the best rewards (lowest latency) based on not only the current query but also the query execution log.
With this, we could capture meaningful context information and query correlations.
We further introduce the hyper-network and adaptor layer inside each transformer block.
The hyper-network accepts inputs of the query and decides which adapter layer this query should use.
If this query is determined as an unseen query, the hyper-network will generate a new adapter with generated initialized parameters to facilitate fast training.
If this is a known query, hyper-network directly uses the existing one.

\subsection{Fast-adaptive learned concurrency control}
\label{subsec:cc}


Concurrency control is crucial in transaction processing to ensure ACID properties~\cite{DBLP:books/mk/GrayR93} and achieve optimal performance. 
Traditional concurrency control algorithms are generally rule-based. 
For instance, two-phase locking (2PL) relies on lock compatibility to order transactions, while optimistic concurrency control (OCC) depends on order validation techniques.
Given the diversity of real-world workloads, a fixed rule-based approach often does not work well across all scenarios.
This limitation has triggered the development of hybrid and learning-based concurrency control algorithms~\cite{DBLP:conf/osdi/WangDWCW0021,DBLP:conf/sigmod/SuCDAX17,DBLP:conf/usenix/TangE18}.
These algorithms typically analyze specific workloads, either manually or automatically, to select the most suitable concurrency control algorithms for different types of transactions.
While they outperform traditional methods in certain scenarios, they struggle to adapt to the dynamic workloads in real-world databases. Consequently, they may not consistently yield optimal concurrency control as workloads and data distribution change.


We aim to design a learned concurrency control algorithm that can continually adapt to changing workloads, ensuring both efficiency and practicality.
Unlike existing approaches, we do not focus on learning from specific workloads. 
Instead, we aim to develop a general concurrency control strategy that reflects workload shifts through system statistics, allowing for dynamic adaptation. 
This means that our concurrency control strategy continuously updates based on system behavior to accommodate workload changes.
However, achieving this objective presents two key challenges. 
First, the system must be modeled precisely to detect workload shifts in real-time. 
Second, upon adaptation, the model update must be as fast and accurate as possible to ensure a rapid response to changing workloads.

To address these challenges, we propose an online learned concurrency control algorithm that ensures quick adaptation to dynamic workloads. 
We first design a system representation by carefully selecting critical factors, such as throughput and average lock waiting time, to represent system conditions accurately. 
Based on these factors, we then establish a general concurrency control strategy that maps specific system conditions to corresponding actions.
For each transaction, this strategy determines the appropriate concurrency control action. 
For instance, when a write operation is performed on high-contention data items, we require a lock immediately to avoid wasted context switching overhead, while for a read operation on low-contention data, we avoid locking to enhance performance.

By monitoring these factors, we can detect workload changes through their impact on system conditions. 
For example, if the lock waiting time suddenly spikes, it indicates that the system is under high load.
In this case, we need to trigger the concurrency control strategy adaption.
We then propose a fast concurrency control adaptation model based on online Reinforcement Learning (RL).
To ensure the strategy can continuously update with changing workloads, we employ a two-phase adjustment algorithm following FRP. 
In the filtering phase, we generate several improved strategies using an evolutionary algorithm (EA) and test them over a specific timeframe to identify the top strategy. 
In the subsequent refinement phase, we use reward-based feedback to further refine and improve the selected strategy.

\section{Privacy-preserving and trusted \dbname} \label{sec:secure}

In this section, we present the architecture of \dbname from the privacy and trustworthiness perspective, detail in-database federated learning, and discuss the online recovery technique to correct data tampering.

\begin{figure}[t]
\centering
\includegraphics[width=0.83\columnwidth]{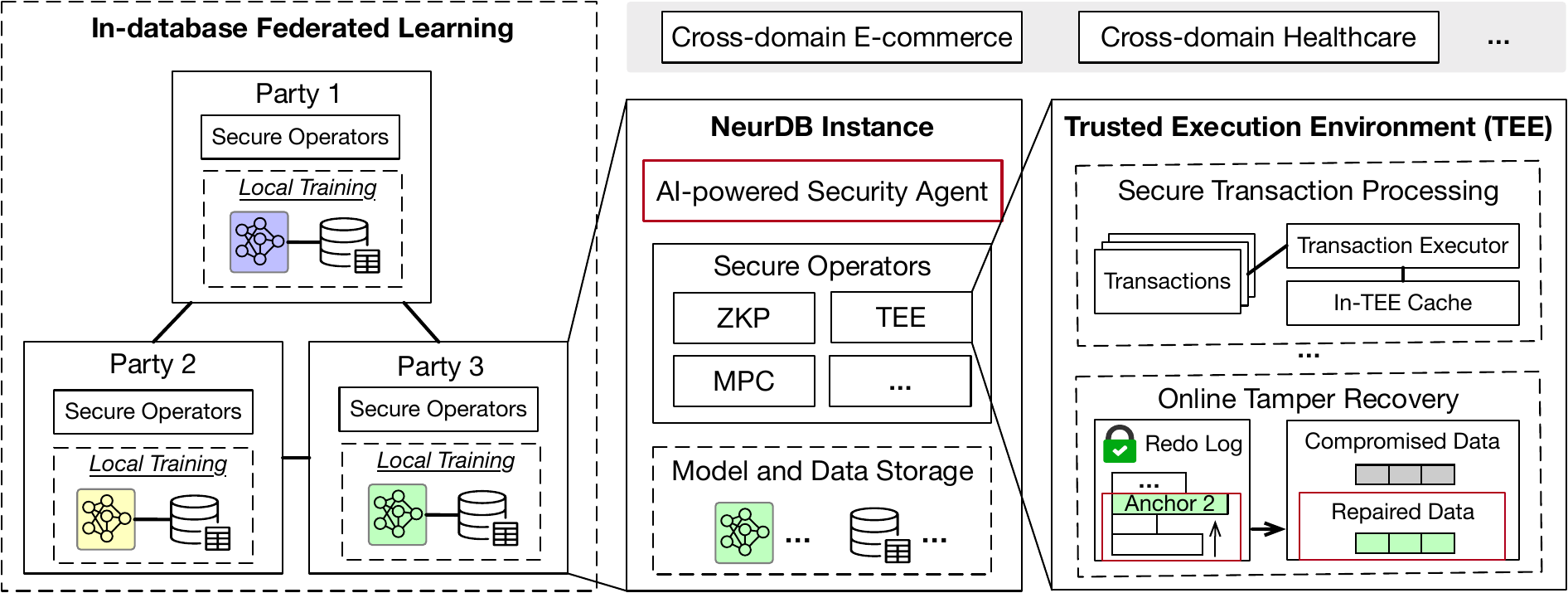}
\caption{Overview of privacy-preserving and trusted \dbname}
\label{fig:overview_security}
\end{figure}

\subsection{Privacy-preserving and trusted architecture overview}
\label{subsec:sec_overview}

With the escalated demand for performing collaborative analytics on sensitive data, data privacy and trustworthiness have become indispensable. For example, different organizations may train machine learning models on their joint databases, where each organization must protect its own data while being able to verify the authenticity and execution correctness of data from others.
Another example is that users may upload data and models to the database systems deployed on the cloud for data analytics; users have to ensure that the data and models are protected and not tampered with by the cloud. 
%
To cope with the various needs of users, we design the privacy-preserving and trusted functions of \dbname based on the following three design goals:
a) the secure functions for data privacy and trustworthiness can be added on demand and easily managed by users, b) the database system must natively support privacy-preserving data collaboration while being adaptive to prevent the malicious behaviors timely, c) the system must be autonomous and able to generate task-specific privacy and trustworthiness policies.

\dbname presents an integral solution of AI$\times$DB to guarantee data privacy and trustworthiness. First, it introduces various built-in secure operators to support hot plugging and user-friendly secure services. Second, it supports in-database federated learning, which enables privacy-preserving and trusted data collaboration. Lastly, \dbname embeds an AI-powered agent to automate privacy-preserving and trusted task generation and policy configuration.
Figure~\ref{fig:overview_security} shows the architecture of our privacy-preserving and trusted \dbname. It consists of a task handler, a coordinator, executors, and an AI-powered security agent. The task handler is responsible for parsing the users' requests and generating execution plans for the secure operators. The coordinator distributes the task to the executors. 
The latter performs the tasks according to the privacy and trustworthiness policies of the operators and returns the results to the coordinator. 
The AI-powered security agent monitors the users' requests, and assists users with the task generation and policy configuration.

To enhance the inherent privacy and trustworthiness of \dbname, we redesign the database's security paradigm by introducing new privacy-preserving and trusted operators.
These operators are designed to trigger autonomously in response to predetermined criteria, thereby removing the necessity for manual intervention. 
Based on our extensive background in federated learning~\cite{DBLP:journals/pvldb/WuX0DLOX023,DBLP:journals/pvldb/WuCXCO20,DBLP:conf/mm/OoiTWWCCGLTWXZZ15,DBLP:journals/zkpfl24}, privacy-preserving data analytics~\cite{DBLP:conf/kdd/BaoGXL23,DBLP:conf/ccs/SunXKYQWY19}, and database security~\cite{DBLP:journals/pvldb/YueDXZCOX23,DBLP:journals/pvldb/WangDLXZCCOR18}, we develop innovative techniques for facilitating the autonomous deployment of these operators. We describe the fundamental operators as follows:

\begin{itemize}[leftmargin=*]
    \item \texttt{ENC(TYPE, TEXT), ENC\_OP(TYPE, MSG, MSG), DEC(TYPE, MSG)}. 
    These operators are responsible for encrypting data, decrypting data, and executing operations on encrypted messages. The \texttt{ENC} operator takes the encryption scheme and plaintext data as input and returns the encrypted message. We support various encryption schemes such as fully homomorphic encryption (FHE) and partial homomorphic encryption (PHE) with additive and multiplicative homomorphism. \texttt{ENC\_OP} takes in encrypted or plaintext messages and calculates the results of the corresponding operations. \texttt{DEC} will decrypt the message back to plaintext with the secret keys.
    \item \texttt{MPC(OP, PARTY.DATA, ...)}. Secure Multi-Party Computation (MPC)
    enables multiple parties to jointly calculate a function using their individual inputs while keeping those inputs private.
    The \texttt{MPC} operator supports multi-party computation by taking as input the operator and data from a number of parties. It returns the final result to the users hiding all the details of the MPC protocol. We support the state-of-the-art additive secret sharing scheme (SSS) in MP-SPDZ.
    \item \texttt{DP(DATA)}. Differential privacy (DP) offers a rigorous privacy guarantee for protecting individuals' sensitive information.
    It ensures that the probability of producing a given output does not depend very much on whether a particular individual is included in the database. 
    The \texttt{DP} operator allows the system to add controlled random noise before releasing the output, e.g., the trained model.
    \item \texttt{ZKP(DATA)}. Zero-knowledge proofs (ZKPs) allow one party (the prover) to prove the truth of a statement to another party (the verifier) without revealing any information other than the statement's truthfulness. The \texttt{ZKP} operator allows users to query the data and verify the correctness of the data and operation with the proofs and operation with the proofs provided by the system.
    \item \texttt{TEE}. \dbname provides options for privacy-preserving and trusted operations with the support of trusted hardware. By specifying \texttt{USING TEE}, all user queries and data will be executed and stored within the trusted execution environment (TEE), such as Intel SGX (Software Guard Extensions), so that other adversaries in the system could not exploit such private information.
    \item \texttt{LEDGER}. \dbname supports distributed ledger schema with the keyword \texttt{LEDGER}. Data stored in \texttt{LEDGER} is immutable, transparent, and verifiable. \texttt{LEDGER} guarantees the correctness of the data, history, and queries. It also tracks the lineage of data. For all queries involving \texttt{LEDGER}, the system will provide integrity proof in addition to the data selected or inserted into the \texttt{LEDGER}.
\end{itemize}

\subsection{In-database federated learning}


To enable privacy-preserving and trusted data collaboration, \dbname supports in-database federated learning leveraging the secure operators defined in Section~\ref{subsec:sec_overview}. In traditional federated learning scenarios, users need to extract the datasets from their databases and preprocess the datasets before sending them to the models for training. The whole process requires multiple rounds of data processing and copying, leading to significant overhead. In contrast, data in \dbname are ready for training due to the built-in secure operators. Data can go through pre-processing and training while being queried.

As shown in Figure~\ref{fig:overview_security}, the training process is distributed across multiple decentralized \dbname instances without needing to move or centralize the data. Each participating instance retains its data locally. Instead of moving data to a central location, the federated learning algorithm enables the model to be trained locally on each database and only the aggregated global model is released. 
The aggregator combines these parameters from all participating models to create an updated, global model. This process of training locally and aggregating globally repeats several times. Each iteration refines the global model further, without ever exposing the individual datasets. 
\dbname supports both horizontal and vertical federated learning to handle various scenarios.

To protect data privacy and ensure trustworthiness for \dbname's in-database federated learning with high efficiency and utility, we design novel protocols based on the built-in secure operators. 
First, we propose a hybrid technique based on MPC and PHE operators. This technique employs PHE as much as possible to facilitate each party's local computations and utilizes MPC when PHE cannot compute the functions. This technique protects the information exchanged among the parties for federated training and improves the efficiency of secure computations. 
Second, we incorporate DP into in-database federated learning to prevent the adversary from inferring the sensitive information of each party's local data. Specifically, we formalize a novel notion of label distributional privacy and propose a label distribution perturbation mechanism, which is a variant of stochastic gradient descent (SGD) that satisfies label distributional privacy, protecting each party's labels during federated training. 
Third, we present a ZKP-based federated learning algorithm that is highly efficient for secure and verifiable data collaboration, which not only protects the privacy of model updates that each party uploads to the server, but also guarantees that the model updates are well-formed such that malicious updates can be detected. Particularly, we devise an optimized ZKP generation and verification technique that significantly reduces the ZKP cost based on probabilistic integrity checks.
Fourth, we introduce an efficient, verifiable distributed ledger using a two-level Merkle-like tree structure to log the training and prediction activities of in-database federated learning, which ensures an immutable and transparent collaboration history with high throughput.



\subsection{Tamper recovery and fault tolerance}

Data tampering is a critical issue that can potentially cause a system to be unavailable.
If data is tampered with, transactions accessing this data need to be blocked to prevent further damage.
Existing blockchain systems based on Byzantine Fault Tolerance (BFT) protocols can mitigate this issue but suffer from significant performance overhead. 
Another line of related research, namely verifiable databases, typically restores the entire system to a previous snapshot, which can lead to the loss of recent updates.
This demonstrates that correct and efficient tampering recovery remains an open problem.



Unlike existing works, we aim to prevent the system from being unavailable due to data tampering without additional overhead on normal transaction processing.
To achieve this, we execute transactions on data that require protection by specifying \texttt{USING TEE}.
All transaction operations are performed in TEE so that if required data is tampered with, it can be detected when it is loaded into TEE.
If data tampering is detected, we employ a log-based tampering recovery mechanism that allows transactions to continue once the tampered data is repaired.
Inspired by commonly used log recovery techniques~\cite{DBLP:conf/sigmod/ArasuEKKMPR17}, we consider replaying all previous operations on the compromised data for correction.
To this end, we enhance the redo log to be secure by encrypting it with a signature-based mechanism.
Specifically, we generate a unique hash for each transaction's log entries, which is 
signed in TEE to ensure any tampering with the log can be detected. 
We then retrieve corresponding logs into the TEE, in which we replay the missing operations to repair the data to its correct state. 
With TEE, the recovery process is shielded from tampering.


Achieving efficient log replay for tampering recovery is not straightforward.
Since traditional redo logs only contain information about changes, rebuilding a corrupted data item requires replaying from the first log entry that created it. This process can lead to lengthy replay times due to the large volume of logs that must be replayed.
We therefore introduce {anchor logs} to facilitate efficient and online recovery. 
In particular, an anchor log is created after every $n$ modification to a specific data item, where $n$ is a tunable parameter.
Differing from standard redo logs, anchor logs record the entire content of the data item.
For example, consider a transaction that updates a data item $x_i$ to $x'_i$, representing the $n$-{th} modification of $x_i$ since the last anchor log.
In response, we create a new anchor log that captures the entire content of $x'_i$.
This allows the recovery process to start from the most recent anchor log, applying subsequent changes incrementally, thereby avoiding replicating logs from the beginning.
After the recovery process, the affected transactions can continue execution with the correct data.

\section{Implementation of \dbname} \label{sec:nextstep}
We shall now discuss the development of \dbname, particularly focusing on its implementation.
\dbname is built on the PostgreSQL codebase, which we significantly enhance with our proposed AI-powered architecture and techniques.
In the implementation, 
we are concerned with
efficient interactions between the database engine and in-database AI models. 
Minimizing interaction with AI models is not straightforward for the following two key reasons.
First, AI models are often implemented in Python, while the database is primarily written in C/C++, which incurs overhead from data copying between different programming languages.
Second, model updates in \dbname are necessary to adapt to dynamic workloads.
However, frequent updates to AI models can be costly if each update requires reloading the models.
To address these issues, we leverage a carefully designed shared memory system to avoid the overhead of data and model transfers.
 In particular, we allocate shared memory accessible in both Python and C/C++ environments. 
Data needed by AI models is directly retrieved into this shared memory, which allows the Python environment to access and extract data without additional copying overhead. 
Similarly, model updates are performed directly in the shared memory, enabling immediate access without additional model loading.
Further, we introduce a pipeline technique to ensure that data retrieval can be performed concurrently with other computationally intensive operations, such as model training~\cite{DBLP:conf/mm/OoiTWWCCGLTWXZZ15,singaeasy,DBLP:conf/mm/WangCDGOTW15}.
To facilitate this, we organize the shared buffer as a circular buffer.
For example, a background thread is employed to regularly check the availability of cache space and load batches of data into the cache, which ensures a continuous supply of data necessary for model training.

We shall next describe the current implementation of AI-friendly operators in \dbname.
To ensure that users can conveniently invoke these operators, we implement them based on user-defined functions (UDFs), allowing operators to be called with SQL statements.
For example, let us consider the model slicing operator.
When activated, the UDF for model slicing performs four key tasks.
First, it identifies the prediction target and determines the appropriate general model.
Second, it customizes the model via the proposed SQL-aware dynamic model slicing technique based on the \textit{filter} conditions in the query and then caches the sliced model in the shared memory.
Third, it retrieves relevant data according to the filter conditions using the server programming interface (SPI) and stores the selected data in shared memory. 
Finally, it performs the model inference with the data and outputs the results.
The development of \dbname has unveiled as many questions as it has resolved.
Further improvement in performance, scalability, and functionalities remain the development goals of \dbname. 
We shall report new results in the near future
and release the code at appropriate time \cite{neurdbwebsite}.

\section{Conclusion} \label{sec:conclusion}

New applications 
are
often catalysts for driving
technological innovations. 
With the immense potential of 
AI on all aspects of data-centric applications, data systems must now seamlessly support AI- and data-driven analytics.
Naturally,
AI-powered applications
require next-generation data systems to 
efficiently support AI applications and fully leverage AI technologies. 
In this paper, we present the conceptual and architectural overview of our proposed AI-powered autonomous data system, called \dbname. 
We also outline the innovative techniques developed for \dbname, and 
envision
its future trajectory
in system implementation and evolution.





\Acknowledgements{
We would like to thank Ergute Bao, Guoyu Hu, Hexiang Pan, and other colleagues for their help and comments.
The work of NUS researchers is partially supported by the Lee Foundation in terms of Beng Chin Ooi's Lee Kong Chian Centennial Professorship fund and NUS Faculty Development Fund.
Gang Chen's work is supported by National Key Research and Development Program of China (2022YFB2703100).
Yuncheng Wu's work is supported by National Key Research and Development Program of China (2023YFB4503600).
Meihui Zhang's work is supported by National Natural Science Foundation of China (62072033).
}



\bibliographystyle{unsrt}
\balance
\bibliography{main}

\end{document}